\documentstyle[12pt,titlepage]{article}

\title{On the Logarithmic Triviality of Scalar Quantum Electrodynamics}

\author{by\\[8pt]  M. Baig and H. Fort\\ Grup de F\'{i}sica Te\`orica\\
Institut de F\'{i}sica
d'Altes Energies\\Universitat Aut\`onoma de Barcelona\\ 08193
Bellaterra (Barcelona) SPAIN\and J. B. Kogut\\ Physics Department\\ 1110
West Green Street\\University of Illinois\\ Urbana, IL  61801-3080 \\ \\
S. Kim and D. K. Sinclair\\ High Energy Physics Division\\
Argonne National Laboratory\\ Argonnne, IL  60439}

\begin{document}


\maketitle

\begin{abstract}

Using finite size scaling and histogram methods we obtain numerical
results from lattice simulations indicating the logarithmic
triviality of scalar quantum electrodynamics, even when the bare
gauge coupling is chosen large.  Simulations of the non-compact
formulation of the lattice abelian Higgs model with fixed length
scalar fields on $L^{4}$ lattices with $L$ ranging from $6$ through
$20$ indicate a line of second order critical points.
Fluctuation-induced first order transitions are ruled out.  Runs of
over ten million sweeps for each $L$ produce specific heat peaks
which grow logarithmically with $L$ and whose critical couplings shift
with $L$ picking out a correlation length exponent of $0.50(5)$
consistent with mean field theory.
This behavior is qualitatively similar to that found in pure
$\lambda\phi^{4}$.

\end{abstract}

\newpage

Do field theories which are strongly coupled at short distances exist
in four dimensions?  This is an important question to answer from both
a purely theoretical and a phenomenological perspective.
Theoretically, one wants to know if the Landau zero$^{1.}$ (complete
screening
of interactions) is universal in non-asymptotically free field
theories in four dimensions, as suggested by perturbation theory.
In less than four dimensions, theories with non-trivial high energy
interactions are commonplace and perturbation theory in bare coupling
parameters is known to be misleading.  For example, four Fermi
theories in dimensions $d$, $2<d<4$, have an ultra-violet stable
fixed point where chiral symmetry is broken, as indicated by $1/N$
expansions while perturbative expansions in the theory's coupling
constant are non-renormalizable.$^{2.}$  Phenomenologically, one wants to
understand the Higgs mechanism in the successful Standard Model and
build theories where the appropriate form of spontaneous symmetry
breaking can occur at short distances.

Issues such as these have rekindled interest in existence questions
for various field theories.  Considerable work on $\lambda\phi^{4}$
theories strongly suggest that this theory becomes free as its cutoff
is removed,$^{3.}$ although a proof of this property remains elusive.  Of
course, $\lambda\phi^{4}$ theories are unrealistically simple so the
phenomenological impact of this program is not great.  In this paper
we shall study scalar electrodynamics at strong gauge couplings with
sufficient numerical resources to make quantitative claims about its
ultraviolet behavior.  We shall see that our numerical results are
consistent with the logarithmic triviality of scalar electrodynamics,
qualitatively similar to pure $\lambda\phi^{4}$.

We begin with a lattice formulation of scalar electrodynamics which
is particularly well suited to answer these physics questions.
Consider the non-compact formulation of the abelian Higgs model with
a fixed length scalar field,$^{4.}$

\begin{equation}
S=\frac{1}{2}\beta\sum_{p}\theta_{p}^{2}-\lambda\sum_{x,\mu}(\phi_{x}^{\ast}
U_{x,\mu}
\phi_{x+\mu}+c.c.)
\end{equation}

\noindent
where p denotes plaquettes, $\theta_{p}$ is the circulation of the
non-compact
gauge field $\theta_{x,\mu}$ around a plaquette, $\beta=1/e^{2}$ and
$\phi_{x}=exp(i\alpha(x))$ is a phase factor at each site.  We choose this
action (the electrodynamics of the planar model) because preliminary work
has
suggested that it has a line of second order transitions,$^{4.}$ because it
does not
require fine tuning and because it is believed to lie in the same
universality class as the ordinary lattice abelian Higgs model with a
conventional, variable length scalar field.$^{5.}$  In Fig. 1 we show the
phase
diagram of the model in the bare parameter space $\beta-\lambda$.  A
preliminary investigation has indicated that the line emanating from the
$\beta\rightarrow\infty$ limit of Fig. 1 is a line of critical points
which
potentially could produce a family of interacting, continuum field
theories.$^{4.}$  Note that in the $\beta\rightarrow\infty$ limit the gauge
field
in Eq. (1) reduces to a pure gauge transformation so the model becomes
the four dimensional planar model which is known to have a second order
phase transition which is trivial, i.e. is described by a free field.  The
non-compact nature of the gauge field is important in Fig. 1---the compact
model has a line of first order transitions and only at the endpoint of
such
a line in the interior of a phase diagram can one hope to have a critical
point where a continuum field theory might exist.$^{4.}$  Since one must
fine tune
bare parameters to find such a point, the compact formulation of the model
is much harder to use for quantitative work.$^{6.}$  The fact that Eq. (1)
uses
fixed length scalar fields avoids another fine tuning---the variable length
scalar field formulation would possess a quadratically divergent bare mass
parameter which would have to be tuned to zero with extraordinary accuracy
to search for critical behavior.  Conventional wisdom based on the
renormalization group states that Eq. (1) should have the same critical
behavior as the fine-tuned variable length model,$^{5.}$ so it again
emerges as
preferable.  Note also that in the naive classical limit where the field
varies smoothly
Eq. (1) reduces to a free massive vector boson.  In the vicinity of the
strong coupling critical point we investigate here, the fields are rapidly
varying on the scale of the lattice spacing and the specific heat scaling
law is not that of a Gaussian model.

First consider the measurements of the internal energies,

\begin{equation}
E_{\gamma}=\frac{1}{2}<\sum_{p} \theta_{p} ^{2}> ,\hspace{.5in} E_{h} =
<\sum_{x,\mu}
\phi_{x} ^{\ast} U_{x,\mu} \phi_{x+\mu} + c.c.>
\end{equation}

\noindent
and their associated specific heats $C_{\gamma} = \partial
E_{\gamma}/\partial
\beta$, and $C_{h} = \partial E_{h}/\partial\lambda$.   Non-analytic
behavior in the specific heats at critical
couplings can be used to find and classify phase transitions.  On a $L^{4}$
lattice
the size dependence of a generic specific heat at a second order critical
point
should scale as,$^{7.}$

\begin{equation}
C_{max} (L) \sim L^{\alpha/\nu}
\end{equation}

\noindent
where $\alpha$ and $\nu$ are the usual specific heat and correlation
length
critical indices, respectively.  Here $C_{max}$ denotes the peak of the
specific
heat.  A measurement of the index $\nu$ can be made from the size
dependence of
the position of the peak.  In a model which depends on just one coupling,
call it
$g$, then$^{7.}$

\begin{equation}
g_{c} (L) - g_{c} \sim L^{-1/\nu}
\end{equation}

\noindent
where $g_{c} (L)$ is the coupling where $C_{max} (L)$ occurs and $g_{c}$ is
its
$L\rightarrow\infty$ thermodynamic limit.  The scaling laws Eq. (3) and (4)
characterize a critical point with powerlaw singularities.  This is a
possible
behavior for scalar electrodynamics, but there is also the possibility
suggested by
perturbation theory, that the theory is logarithmically trivial.  Consider
$\lambda
\phi^{4}$ as the simplest, well-studied theory which apparently has this
behavior.
In this case the theory becomes trivial at a logarithmic rate as the
theory's momentum
space cutoff $\Lambda$ is taken to infinity.  Then the scaling laws of
Eq. (3) and
(4) become,$^{8,9}$

\begin{equation}
C_{max}(L)\sim({\ell}nL)^{p}
\end{equation}

\noindent
and

\begin{equation}
g_{c}(L)-g_{c}\sim\frac{1}{L^{2}({\ell}nL)^{q}}
\end{equation}

\noindent
where $p$ and $q$ are powers predictable in one-loop perturbation theory
($p=\frac{1}{3}$
and $q=\frac{1}{6}$ in $\lambda\phi^{4}$).  Note the differences between
these scaling
laws and those of the usual Gaussian model, obtained from Eq. (3) and (4)
setting
$\alpha=0$ and $\nu=.5$:  in the Gaussian model the specific heat should
saturate as $L$
grows, and the position of the peaks should approach a limiting value at a
rate $L^{-2}$.

It is particularly interesting in scalar electrodynamics to consider a
large value
of the bare (lattice) gauge coupling to see if that can induce non-trivial
interactions which survive in the continuum limit.  So, we ran extensive
simulations on lattices ranging from $6^{4}$ through $20^{4}$ at
$e^{2}=5.0$ and
searched in parameter space $(\beta,\lambda)$ for peaks in $C_{\gamma}$ and
$C_{h}$.  We used histogram methods$^{10,11}$ to do this as efficiently as
possible.
For example, on a $6^{4}$ lattice at $\beta=.2000$ and $\lambda=.2350$ we
found a specific
heat peak near $\lambda_{c}(6)\approx.2382$ from the histogram method.  The
$\lambda$
value in the lattice action was then tuned to .2382 and additional
simulations and
histograms produced specific heats, found from the variances of
$E_{\gamma}$ and
$E_{h}$ measurements, at a $\lambda_{c}$ very close to .2382.  Using this
strategy,
measurements of $\lambda_{c}(L),C_{\gamma}(L)$ and $C_{h}(L)$ could be made
without
relying on any extrapolation methods.  We thus avoided systematic errors,
although
critical slowing down on the larger lattices limited our statistical
accuracy.  In Table 1
we show a subset of our results that will be analyzed and discussed here.
A more
complete discussion with additional measurements and analysis will appear
elsewhere.  The columns labeled $\lambda_{c}(L),C_{\gamma}^{max}(L)$ and
$C_{h}^{max}(L)$ in Table 1 need no further explanation except to note that
the
error bars were obtained with standard binning procedures which account for
the
correlations in the data sets produced by Monte Carlo programs.  The Monte
Carlo
procedure used here was a standard multi-hit Metropolis for the non-compact
gauge
degrees of freedom and an over-relaxed plus Metropolis
algorithm$^{12.}$ for the
compact
matter field.  Over-relaxation reduced the correlation times in the
algorithm by
typically a factor of 2--3.  Accuracy and good estimates of error bars are
essential in a quantitative study such as this.  Unfortunately, cluster and
acceleration algorithms have not been developed for gauge theories, so very
high
statistics of our over-relaxed Metropolis algorithm were essential---tens
of
millions of sweeps were accumulated for each lattice size as listed in
column 7.
Specific heats were measured as the fluctuations in internal energy
measurements
$(C_{h}=(<E_{h}^{2}>-<E_{h}>^{2})/4L^{4}$, etc.), and very high statistics
and many $L$ values
are needed to distinguish between logarithmic triviality (Eq. 5) and
powerlaw
behavior (Eq. 3).  The other entries in the Table, $K_{\gamma}(L)$ and
$K_{h}(L)$, are
the Binder Cumulants (Kurtosis)$^{13.}$ for each internal energy.  At a
continuous
phase  transition each Kurtosis should approach $2/3$ with finite size
corrections
scaling as  $1/L^{4}$.  The Kurtosis is a useful probe into the order of a
phase
transition, although  an examination of the internal energy and specific
heat
histograms are often just as  valuable. Since the order of the transitions
in lattice
and continuum scalar  electrodynamics are controversial, we studied these
quantities
with some care.

As a warmup to the full theory, we checked that our techniques are able to
reproduce known results.  For example, when $\beta\rightarrow\infty$ Eq.
(1)
reduces to the four dimensional planar spin model which should have an
order-disorder transition as a function of $\lambda$ that is
described by
mean field theory $(\alpha=0, \nu=.5, etc.)$ and produce a free massless
field in the continuum limit.  We measured the specific heat at the
transition for
$L=6$, 8, 10, 12 and 14, and found peak values 20.47(3), 22.80(5),
24.35(9), 25.38(9)
and 26.24(9), respectively.  One million sweeps of our code, tailored for
$\beta=\infty$, were
run in each case.  We note that the specific heat peaks do grow with $L$,
but a three
parameter fit of the form $aL^{\rho}+b$ is excellent (confidence level =
92\%)
producing
$\rho=-.67(11)$, $a=-44(3)$ and $b=33.7(1.7)$.  So, the specific heat is
predicted to
saturate producing a critical index $\alpha=0$, and the growth of the peak
heights seen in
the simulation is a subdominant effect. Other fits to the data such as
$a {\ell}n^{\rho}L+b$ were not stable and over a wide range of fitting
parameters produced confidence levels less than a few percent.
Certainly much more exacting
studies of this model
could be made (cluster algorithms), but we are testing here just the
simulation and analysis
technology available to the gauge model.

Consider the Kurtosis $K_{\gamma} (L)$, the specific heat $C_{\gamma}^{max}
(L)$
and the critical coupling $\lambda_{c}(L)$ of scalar electrodynamics.  As
stated
above, we set the lattice (bare) gauge coupling to $e^{2} = 5.0$ and then
used
simulations, enhanced by histogram methods, to locate the transition line
in
Fig. 1.  The Kurtosis $K_{\gamma} (L)$ is plotted against $10^{6}/L^{4}$ in
Fig. 2.
The size of the symbols include the error bars, but clearly the curve
favors a second order
transition.  A three parameter fit to the $L=12$, 14, 16, 18 and 20 data
using the form
$K_{\gamma}(L)=aL^{\rho}+b$ is excellent (confidence level = 98\%)
predicting
$\rho=-4.1(4)$ and $K_{p}(\infty)=.666665(2)$.  The hypothesis of a line of
second order
transitions in Fig. 1 appears to be very firm, with no evidence for a
fluctuation-induced
first order transition.  An analysis of $K_{h} (L)$ gives the same
conclusion with
somewhat larger error bars.  In Fig. 3 we plot our $C_{\gamma}^{max} (L)$
data vs. $L$.
We attempted powerlaw as well as logarithmic finite size scaling
hypotheses.  The powerlaw
hypothesis did \underline{not} produce a stable fit for any reasonable
range of
parameters.  However, logarithmic fits were quite good.  The hypothesis
$C_{\gamma}^{max}
(L) = a {\ell}n^{\rho}L+b$ for $L=8$,10,12,14,16,18 and 20 fit with a
confidence level
$=~90\%$ producing the estimate $\rho=1.4(2)$.  If we considered the range
$L=8-18$, the same fitting form predicted $\rho = 1.5 (3)$ with confidence
level
$=84\%$, and if the range $L=10-20$ were taken we found $\rho=1.4 (5)$ with
confidence level $=78\%$.  The solid line in Fig. 3 is the $L=8-20$ fit.
An analysis
of $C_{h}^{max} (L)$ gave consistent results---the same logarithmic
dependence should
be found in either specific heat---and powerlaw fits to $C_{h}^{max}(L)$
were
also ruled out.  In particular, a fit of the form
$C_{h}^{max}(L)=a{\ell}n^{\rho}L+b$  for $L=8-18$ gave $\rho = 0.9 (3)$
with
confidence level $=82\%$ and for $L=8-20$ gave  $\rho=1.0(2)$ with
confidence level
$=85\%$.  Finally, in Fig. 4 we show $\lambda_{c}(L)$ vs. $10^{4}/L^{2}$.
The error bars
again fall within the symbols in the figure.  The data is clearly
compatible with the
correlation length index $\nu = 0.5$ expected of a theory which is free in
the continuum
limit.  In the case of $\lambda \phi^{4}$ it has proven possible to find
the  logarithm of
Eq. (6) under the dominant $L^{-2}$ behavior by using special
techniques.$^{14.}$   We
do not quite have the accuracy to do that here:  a powerlaw fit to
$\lambda_{c}
(L)=\lambda_{c}+a/L^{1/\nu}$ using $L = 12-20$ predicts $1/\nu = 2.0 (1)$,
$\lambda_{c}=.22825(8)$ with confidence level $=92\%$ and using $L=14-20$
predicts
$1/\nu=1.9(3)$, $\lambda_{c}=.2282 (2)$ with confidence level $=97\%$.

One of the motivations for this study was the recent finding that the
chiral symmetry
breaking transition in non-compact lattice electrodynamics with dynamical
fermions is
\underline{not} described by a logarithmically trivial model.$^{15.}$
Powerlaw
critical behavior has been found with non-trivial critical indices
satisfying
hyperscaling.  The
present negative result for scalar electrodynamics suggests that the chiral
nature of
the transition for fermionic electrodynamics is an essential ingredient for
its
non-triviality.  It remains to be seen, however, if the chiral transition
found on
the lattice produces an interesting continuum field theory.

In conclusion, our numerical results support the notion that scalar
electrodynamics is a logarithmically trivial theory.  We suspect that this
result
could be made even firmer by additional simulation studies which use more
sophisticated techniques such as renormalization groups
transformations$^{5.}$ or
partition function methods.$^{14.}$  Since we did not wish to bias our
study
toward logarithmic triviality, we did not pursue special methods which
require
additional theoretical input in order to be quantitative.  Certainly our
concentration on a line of fixed electric charge in the entire phase
diagram
should be
relaxed.  Hopefully, accelerated Monte Carlo algorithms could be developed
for
scalar electrodynamics so that larger systems could be simulated with
better
control.  A word of warning for the ambitious---the standard CRAY random
number
generator RANF which uses the linear congruent
algorithm with modulus $2^{48}$ proved inadequate
for
lattices whose linear dimension was a power of 2.  Presumably this occurred
because for strides of length $2^{N}$ the period of RANF is reduced from
$2^{46}$
to $2^{(46-N)}$ and the well-known correlations in such generators are
expected to have maximum effect if the distribution is sampled with a
period of $2^{N}$.The simplicity of lattice scalar electrodynamics also
makes it
more susceptible to the correlations in random number generators than other
models.  We discovered this problem when our $16^{4}$ simulations were
unstable and
after considerable investigative work we isolated the problem in the random
number
generator.  We cured the problem by adding extra calls to RANF to
avoid strides of length $2^{N}$.  Problems with generally accepted random
number
generators have been studied systematically in ref. (16).

\section*{Acknowledgement}
The simulation done here used the CRAY C90's at PSC and NERSC.  We thank
these
centers for friendly user access.  Several thousand cpu hours were needed
to
accumulate the statistics listed in Table 1. J.B.K. is supported in part
by the National Science Foundation grant NSF PHY92-00148. D.K.S. and S.K.
are supprted by DOE contract W-31-109-ENG-38. M.B. and H.F. acknowledge
the support of CESCA and CICYT.

\section*{References}
\begin{enumerate}

\item L. D. Landau and I. Ya. Pomeranchuk, Dokl. Akad. Nauk. {\bf 102}, 489
(1955).

\item K. G. Wilson, Phys. Rev. {\bf D17}, 2911 (1973).  B. Rosenstein, B.
J. Warr and
S. H. Park, Phys. Rep. {\bf 205}, 59 (1991).  S. Hands, A. Kocic and J.
Kogut, Phys.
Lett. {\bf B273}, 111 (1991).

\item C. Aragao de Carvalho, S. Caracciolo and J. Frohlich, Nucl. Phys.
{\bf B215}
[FS7], 209 (1983) and reference therein.

\item M. Baig, E. Dagotto, J. Kogut and A. Moreo, Phys. Lett. {\bf B242},
444 (1990).

\item D. Callaway and R. Petronzio, Nucl. Phys. {\bf277B}, 50 (1980).

\item J. L. Alonso et al., Zaragoza preprint, Sep 25, 1992.

\item M. N. Barber, in \underline{Phase Transitions and Critical
Phenomena}, Vol.
VIII, eds. C. Domb and J. Lebowitz (Academic Press, New York:  1983).

\item E. Brezin, J. Physique {\bf 43}, 15 (1982).

\item J. Rudnick, H. Guo and D. Jasnow, J. Stat. Phys. {\bf 41} 353 (1985).

\item M. Falcioni, E. Marinari, M. L. Paciello, G. Parisi and B. Taglienti,
Phys.
Lett. {\bf 108B} 331 (1982).

\item A. M. Ferrenberg and R. H. Swendsen, Phys. Rev. Lett. {\bf 61}, 2635
(1988).

\item M. Creutz, Phys. Rev. {\bf D36}, 515 (1987).

\item K. Binder, M. Challa and D. Landau, Phys. Rev. {\bf B34}, 1841
(1986).

\item R. Kenna and C. B. Lang, Phys. Lett. {\bf 264B}, 396 (1991).

\item A. Kocic, J. Kogut and K. C. Wang, ILL-(TH)-92\#17 (to appear in
Nucl. Phys.).

\item A. M. Ferrenberg, D. Landau and Y. J. Wong, Phys. Rev. Lett. {\bf
69}, 3382
(1992).

\end{enumerate}

\section*{Figure Captions}
\begin{enumerate}

\item The phase diagram of non-compact scalar electrodynamics.

\item The Kurtosis $K_{\gamma}(L)$ vs. $10^{6}/L^{4}$.

\item The specific heat peaks $C_{\gamma}^{max}(L)$ vs. $L$.  The solid
line is the
logarithmic fit discussed in the text.

\item The critical coupling $\lambda_{c}(L)$ vs. $L^{-2}$.

\end{enumerate}

\begin{table}[hbt]
\caption{Measurements on non-compact lattice scalar electrodynamics.}
\footnotesize
\vspace{12pt}
\begin{tabular}{lllllll}
L & $\lambda_c(L)$ & $C_{h}^{max}(L)$ & $K_{h}(L)$ & $C_{\gamma}^{max}(L)$
&
$K_{\gamma}(L)$ & Sweeps(millions) \\
 6 & .23815(1)      & 13.81(2)          & .657668(9)  & 7.965(9)       &
.665784(2) & 40\\
 8 & .23375(3)      & 15.83(2)          & .662954(5)  & 8.083(3)       &
.666374(1) & 60\\
10 & ..23173(1)      & 17.23(4)          & .664892(4)  & 8.285(6)       &
.666544(1) & 60\\
12 & .23070(1)      & 18.43(7)          & .665713(4)  & 8.457(9)       &
.666606(1) & 30\\
14 & .23004(1)      & 19.38(9)          & .666110(3)  & 8.594(15)      &
.666633(1) & 20\\
16 & .22962(1)      & 20.25(13)         & .666319(2)  & 8.747(17)      &
.666647(1) & 12\\
18 & .22933(1)      & 20.85(15)         & .666441(2)  & 8.863(26)      &
.666654(1) & 12\\
20 & .22912(1)      & 21.76(20)         & .666510(2)  & 8.956(20)      &
.666658(1) & 10\\
\end{tabular}
\end{table}

\end{document}